\documentclass[aps,prl,preprint]{revtex4-1}
\usepackage{blindtext}
\usepackage{graphicx}

\begin{document}
\title{Heat and particle flux detachment with stable plasma conditions \\ in the Wendelstein 7-X stellarator fusion experiment}
\author{Marcin Jakubowski}
\author{Ralf K\"onig}
\author{Oliver Schmitz $^\dagger$}
\author{Yuhe Feng}
\author{Maciej Krychowiak}
\author{Matthias Otte}
\author{Felix Reimold}
\author{Andreas Dinklage}
\author{Peter Drewelow}
\author{Florian Effenberg $^{\dagger \dagger}$}
\author{Yu Gao}
\author{Holger Niemann}
\author{Georg Schlisio}
\author{Andrea Pavone}
\author{Thomas Sunn Pedersen}
\author{Uwe Wenzel}
\author{Daihong Zhang}
\author{Sebastijan Brezinsek $\ddagger$}
\author{Sergey Bozhenkov}
\author{Kai Jakob Brunner}
\author{Daniel Carralero*}
\author{Ken Hammond $^{\dagger \dagger}$}
\author{Golo Fuchert}
\author{Jens Knauer}
\author{Andreas Langenberg}
\author{Heinrich Laqua}
\author{Stefan Marsen}
\author{Novimir Pablant $^{\dagger \dagger}$}
\author{Kian Rahbarnia}
\author{Lukas Rudichshauser}
\author{Torsten Stange}
\author{Glen Wurden**}
\author{W7-X team}
\affiliation{Max-Planck-Institut f\"ur Plasmaphysik, Greifswald, Germany}
\affiliation{$^\dagger$ University of Wisconsin, Madison, USA}
\affiliation{$^{\dagger \dagger}$ Princeton Plasma Physics Laboratory, Princeton, USA}
\affiliation{$\ddagger$ Forschungszentrum J\"ulich GmbH, IEK-4, J\"ulich, Germany}
\affiliation{* CIEMAT, Madrid, Spain}
\affiliation{** Los Alamos National Laboratory, Los Alamos, USA}

\begin{abstract}
The reduction of particle and heat fluxes to plasma-facing components is critical for achieving stable conditions for both the plasma and the plasma material interface in magnetic confinement fusion experiments. A stable and reproducible plasma state in which the heat flux is almost completely removed from the material surfaces was discovered recently in the Wendelstein 7-X stellarator experiment. At the same time also particle fluxes are reduced such that material erosion can be mitigated. Sufficient neutral pressure was reached to maintain stable particle exhaust for density control in this plasma state. This regime could be maintained for up to 28 seconds with a minimal feedback control.  
\end{abstract}

\maketitle
\section{Introduction}
\begin{figure}
	\centering
	\includegraphics[width=0.65\linewidth]{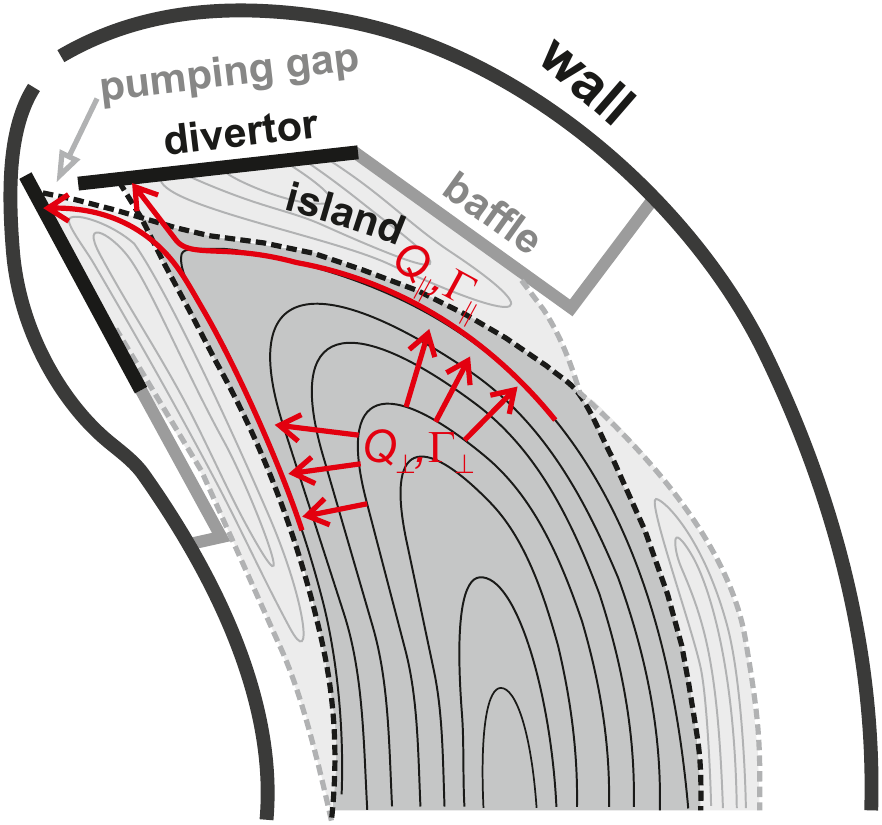}
	\caption{Sketch showing three out of five islands forming the island divertor at Wendelstein 7-X. Black curves indicate cross section of plasma vessel and divertor target plates intersecting magnetic islands. Red arrows indicate heat and particle exhaust channels.}
	\label{fig:poincare}
\end{figure}

The creation of an interface between a high-performance fusion plasma (plasma density of order of $10^{20}$ m$^{-3}$,  plasma temperatures of order of a few keVs, i.e. $10^8$ K) and material surfaces is a grand challenge in fusion energy research. This is especially true for tokamaks, which are characterized by narrow heat exhaust channel at the plasma edge \cite{Eich2013}. A plasma--material interaction, however, is unavoidable because plasma energy confinement is imperfect and the helium as fusion product needs to be exhausted. For steady state plasma conditions in future reactor, heat fluxes in the order of $5$ MW m$^{-2}$ \cite{You2016} and particle fluxes of $>10^{23}$ m$^{-2}$ s$^{-1}$ \cite{Pitts2019} are expected. At the same time the sufficient particle exhaust needs to be assured to allow for efficient control of plasma particle reservoir and helium exhaust. One elegant way to cope with these unprecedented particle and energy fluxes onto material surfaces is to dissipate heat in the plasma before reaching these surfaces, such that heat flux becomes marginal and particle fluxes become strongly reduced. This state is called detachment and it is characterized by significant power flux reduction and plasma pressure losses along a magnetic field line. Detachment results from an interplay of a variety of non-linear physics processes governing interaction between the neutral particles released from the material surfaces and steep density and temperature gradients in the plasma in front of the surface \cite{Stangeby2018}. In the recently started Wendelstein 7-X (W7-X) stellarator experiment, a magnetic island structure (FIG. \ref{fig:poincare}) in front of dedicated surfaces areas is used to define the plasma-material interface. This so-called island divertor geometry allowed the achievement of a stable detachment regime. 

\begin{figure*}[t!]
	\centering
	\includegraphics[width=0.9\linewidth]{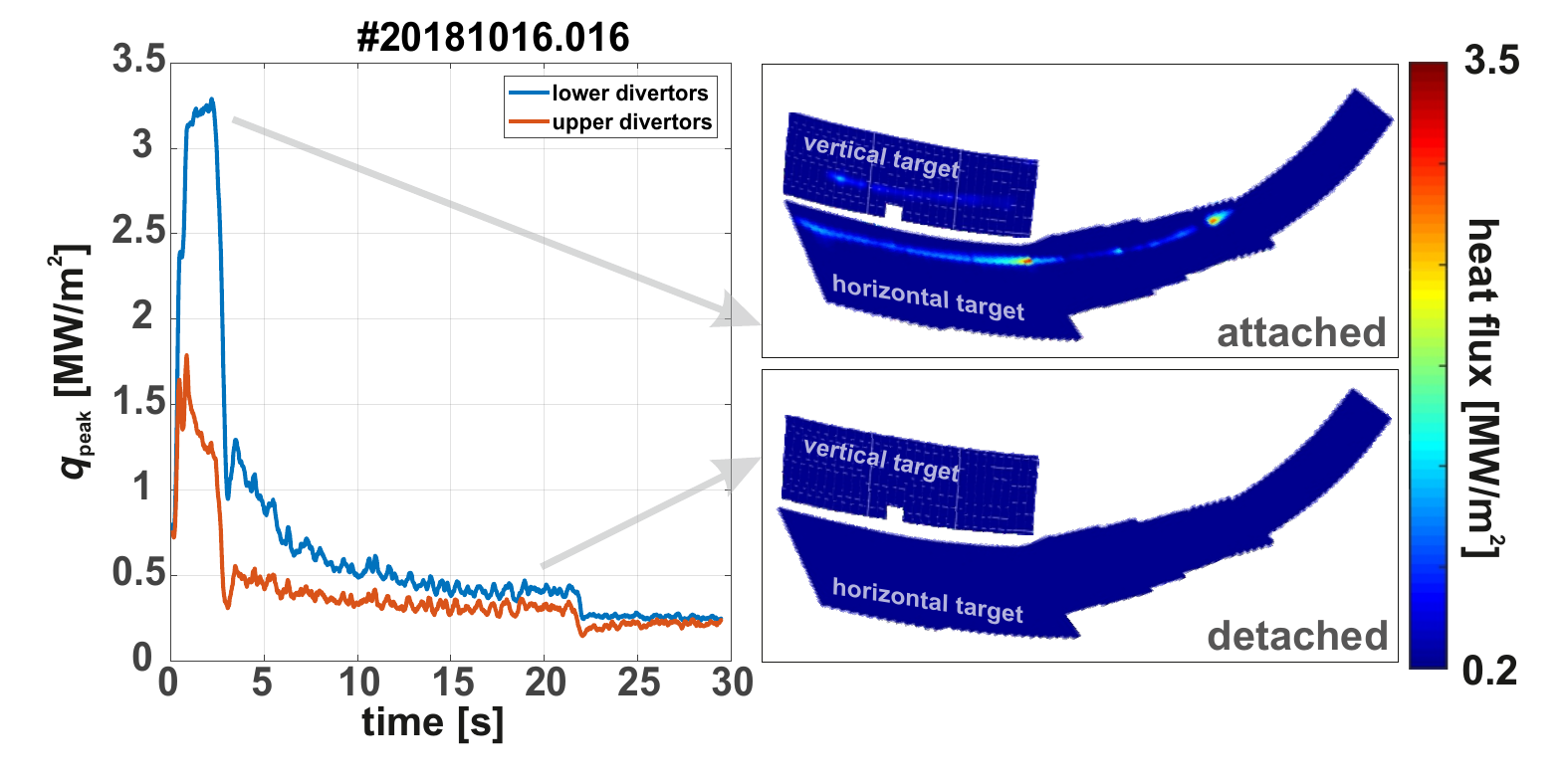}
	\caption{Evolution of heat flux during transition to detachment. (left) Time trace of peak heat flux; (right) snapshot of averaged heat flux distribution at $t = 2$~s and $t = 20$~s).}
	\label{fig:heatflux181016016}
\end{figure*} 
Wendelstein 7-X is the largest advanced stellarator with a major radius of 5.5~m and an average minor radius of ca. 0.5~m. Utilizing superconducting coils a magnetic field of 2.5~T is achieved \cite{Wolf2017a}. Experiments discussed here used electron cyclotron resonance heating (ECRH) to inject power into the plasma. One of the main goals of W7-X is to demonstrate that stellarators can be operated in steady state with a reactor relevant concept of heat and particle exhaust. This is realized with help of so called {\em island divertor}, which consists of 10 discrete divertor units mounted in 5 almost identical modules of the plasma vessel \cite{Pedersen2019} following first realization at Wendelstein 7-AS \cite{Grigull2003a}. Each of the divertors utilizes poloidally extended magnetic islands at the plasma boundary to direct heat and particle towards the divertor target plates from the plasma separatrix (see FIG. \ref{fig:poincare}). Plasma particles are guided towards the targets where they are re-released as recycled neutrals. A fraction of neutrals is guided through a pumping gap between divertor target plates to a sub-divertor volume and from there pumped away by turbo-molecular pumps.    

We will show that a stable state with virtually vanishing heat fluxes, reduced particle fluxes and still sufficient neutral pressures for good particle exhaust and stable density control at Wendelstein 7-X can be reached with minimal control needs (i.e. by controlling plasma density) in a reliable fashion and held constant for up to $28$ s, well in excess of a hundred energy confinement times and time required to saturate the first wall neutral particle reservoir.\\

\section{Stable heat flux detachment with reduced particle flux}

A stably detached divertor plasma state has been achieved during the most recent experimental campaign ($2017-2018$) by raising the line integrated density of the plasma to $1.1\cdot 10^{20}$~m$^2$, where it was kept constant with a  feedback control. This is shown in FIG. \ref{fig:heatflux181016016}.  Here, the peak heat flux $q_{\rm peak}$ to the divertor target is shown on the left vs. time. The red trace shows an average for all upper modules and the blue trace for all lower modules.  Due to edge particle drifts at Wendelstein 7-X \cite{Hammond2019}, power loads to lower divertors are higher than to the upper ones (in plasmas with nominal B-field direction) in the attached stated. In the fully attached state lower divertors almost reach 3.5 MW/m$^2$ at $t \approx 2.5$~s, which decreases in fully detached state (for $t > 3$~s) below 0.5 MW/m$^2$. Both averaged peak heat flux values reach this level and hence the up/down asymmetry of the divertor units vanishes during detachment \cite{Hammond2019}. Also it is important to note that the drop in the peak heat flux after 22 seconds is caused by a drop-out of one of the gyrotrons producing ECRH power, nevertheless plasma remained stable in the detached state, which shows the robustness of this plasma state. Comparison of the averaged heat flux distribution (right side of FIG. 2) to the lower divertors shows two strike lines in the attached stated, which virtually disappear in the detachment phase.

\begin{figure}
	\centering
	\includegraphics[width=1.0\linewidth]{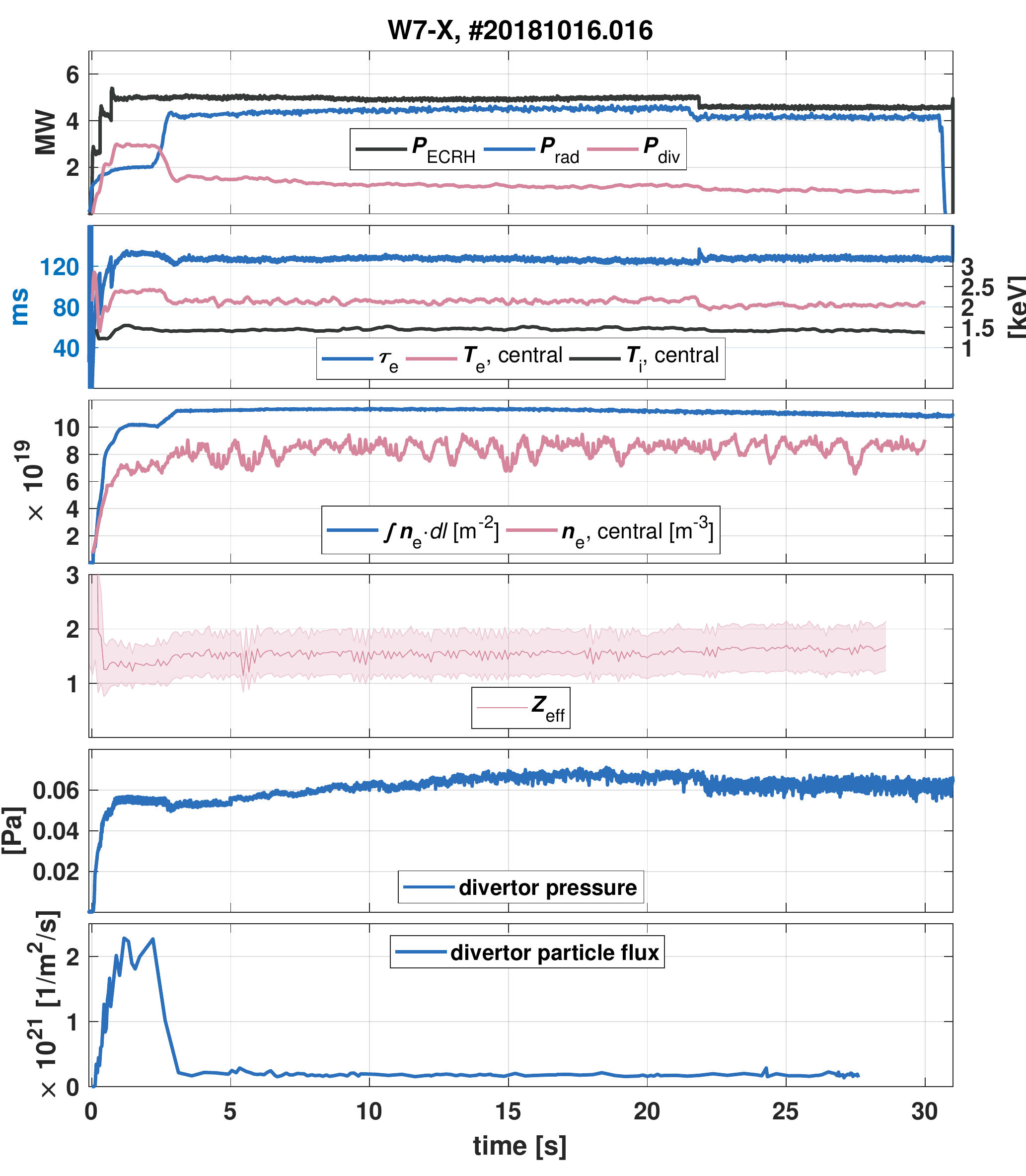}
	\caption{Main parameters of the W7-X discharge \#20181016.016. From top: total heating power ($P_{\rm ECRH}$) delivered by the Electron Cyclotron Resonance Heating, total divertor power ($P_{\rm div}$) and total radiated power ($P_{\rm rad}$); energy confinement time ($\tau_e$), central ion ($T_{\rm ion}$) and electron ($T_{\rm e}$) temperature; line integrated density ($\int n_{\rm e}\cdot dl$) and central electron density ($n_{\rm e}$); effective charge state of plasma -- $Z_{\rm eff}$; divertor pressure measured at the pumping gap and particle flux measured onto one out of ten divertors.}
	\label{fig:Overview16}
\end{figure}

 An overview of the important plasma parameters for this discharge is presented in FIG. \ref{fig:Overview16}. Plasma heating was programmed for $P_{ECRH}=5$ MW using 10 gyrotrons (upper panel) and $30$ seconds and plasma line integrated density $\int n_e \cdot dl$ (third panel from top) was raised to 1.1$\cdot 10^{20}$~[m$^{-2}$] from $t=3$~s at which time complete divertor detachment occurred as seen in the strong reduction of the total power deposited on all divertor modules $P_{div}$ (red trace in upper panel). 
 This plasma state with detached power loading lasted until the pre-programmed end of the discharge. During this detached state also the energy confinement time ($\tau_e \approx 120$~ms) stays almost constant as well as the central electron $T_{\rm e} \approx 2$ keV and ion temperature $T_{\rm i} \approx 1.5$ keV. Together with the reduced heat flux, a significant drop of the particle flux is seen (bottom panel) when the detachment occurs. However, neutral pressure in the divertor $p_{\rm n}$ (second panel from bottom) remains high at $p_{\rm n} \approx 0.07 $ Pa. During the entire detached phase the effective charge of the plasma $Z_{\rm eff}$ remains low at around $1.5$. This means that  no significant increase of impurity concentration occurs with the cooler plasma boundary, i.e. assuming that plasma contains only hydrogen and carbon would lead to ca. 1.8\% concentration of carbon throughout whole discharge). 
 

\section{Onset of detachment}
Detachment in W7-X is realized by increasing plasma radiation, which happens through increasing the plasma density. The main impurity during the experiments at W7-X was carbon that is intrinsically produced by surface material erosion in the divertor. For a given carbon concentration, the amount of power radiated from the impurity is given by $n_{\rm C} \cdot n_{\rm e}^2 \cdot L_z( n_{\rm e},T_{\rm e})$, where $n_{\rm C}$ denotes concentration of carbon impurities, $n_{\rm e}, T_{\rm e}$ -- plasma density and plasma temperature and $L_z$ is the radiated power function. Therefore it is reasonable to expect that lower concentration of carbon is required at high densities to reach the same level of radiation. Numerical studies of detachment at W7-X with EMC3-Eirene \cite{Feng2016} predict that an optimum of plasma radiation ($P_{\rm rad}$) required for effective power dissipation while maintaining sufficient neutral pressure ($p_{n}$) for effective neutral particle pumping is reached for $P_{\rm rad} > 70\%$ of total input power. This is a first experimental verification of the numerical predictions for the island divertor concept, where the interrelation of radiative power losses and the detached divertor conditions manifests itself as predicted in this numerical study.

\begin{figure}[h!]
	\includegraphics[width=\linewidth]{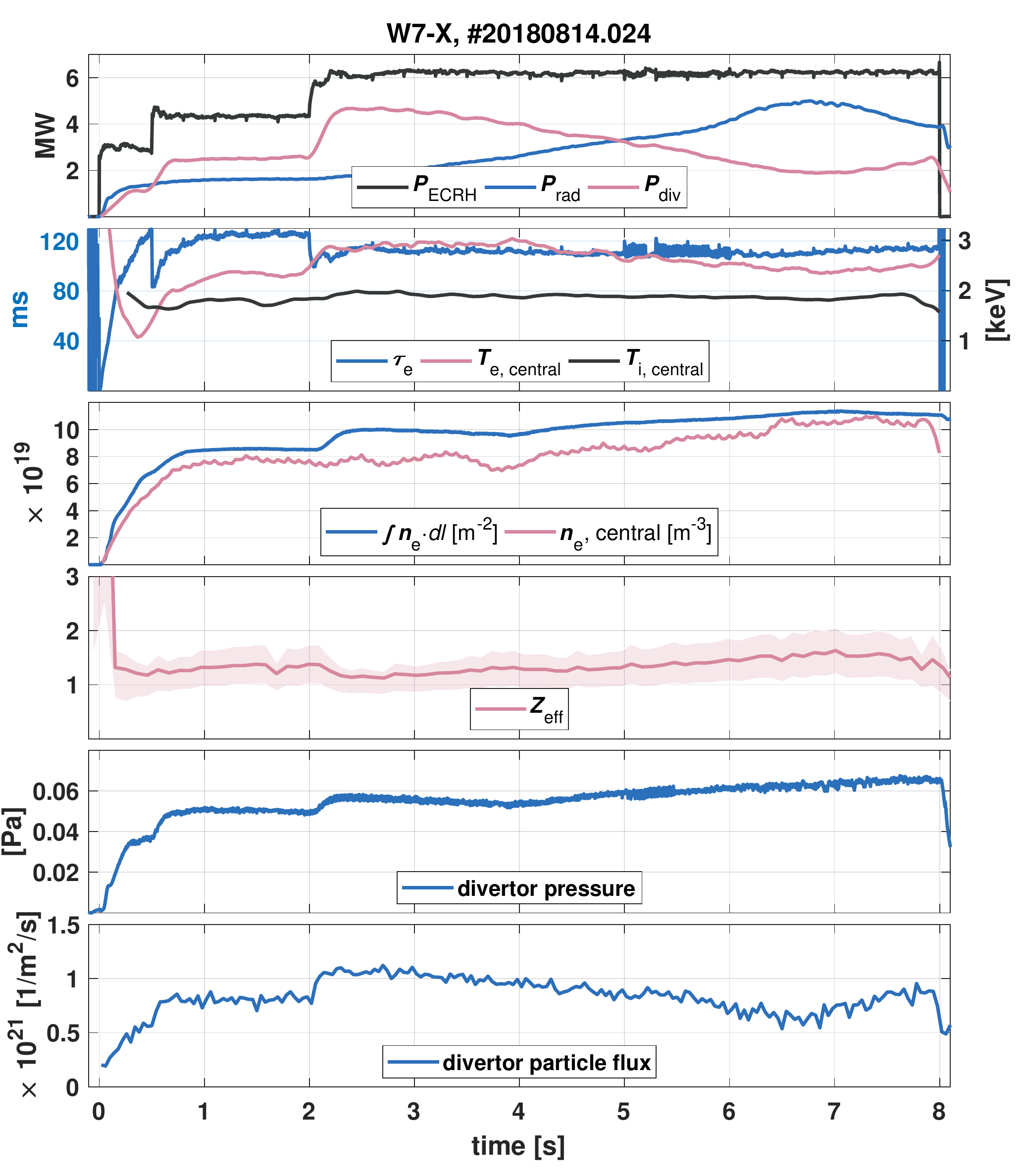}
	\caption{Main parameters of the W7-X discharge \#20180814.024. Description of the time traces is identical to FIG. \ref{fig:Overview16}.}
	\label{fig:overview180814024}
\end{figure}

An example of such  a discharge (\#20180814.024) is presented in Figure \ref{fig:overview180814024}. This discharge used 6 MW of heating power (with ca. 0.97\% of absorbed power) and plasma duration was programmed to about 8 seconds. In this discharge a slow increase of line integrated density $\int n_{\rm e} \cdot dl$ from $9.5\cdot10^{19}$ [m$^{-2}$] to $11.5\cdot10^{19}$ [m$^{-2}$] resulted in a significant increase of the radiated power from ca. 1.8 MW up to ca. 4.5 MW, which means a radiated power fraction increase from $f_{\rm rad} \approx 0.3$ to $f_{\rm rad}\approx 0.8$. During the detached plasma state with high $f_{\rm rad}$, the impurity content of the plasma core is still small as shown by $Z_{\rm eff}=1.5$. As most of the power entering the SOL is dissipated through radiation, total divertor loads (integrated over all 10 divertor units) decrease by more than a factor of 2 from 4.5 MW to 2 MW. A significant fraction (ca. 800 kW) of this deposited power comes from plasma radiation. At $t=7$~s complete thermal detachment is established with simultaneous reduction of the particle flux by a factor of $2$. At the same time relatively high neutral pressure ($p_{\rm n}\approx 0.06$ Pa) is measured in the divertor pumping entrance domain. Only when the radiated power level exceeds about 80\% the divertor pressure starts to drop as predicted by EMC3-Eirene simulations \cite{Feng2016}. Initially $p_{\rm n}$ decreases only slightly, however, when approaching ca. 80\% radiation reduction rate becomes stronger.  Estimates of total pumping rate for detached discharges at ca. $0.6\cdot 10^{21}$ [atoms/s] show that at these densities we reached particle exhaust at the level compatible with steady-state operation.

\begin{figure}
	\includegraphics[width=0.9\linewidth]{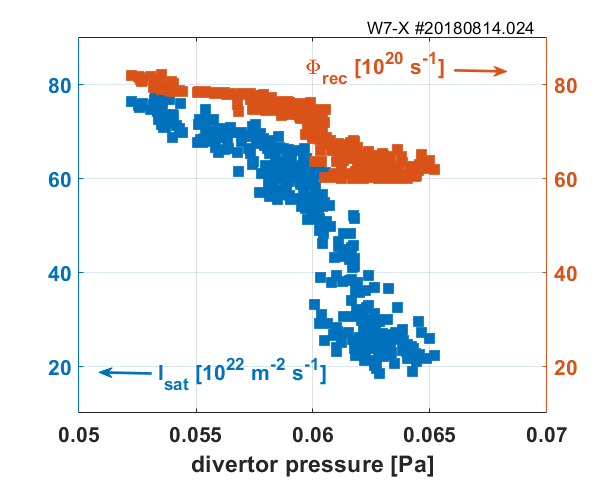}
	\caption{Changes in a total particle flux (red, right abscissa) and ion saturation current measured by a divertor Langmuir probe (blue, left abscissa) versus divertor neutral pressure for discharge \#20180814.024.}
	\label{fig:particleflux}
\end{figure}

FIG.\ref{fig:particleflux} presents the relation between particle flux measured by the cameras equipped with an H$_{\alpha}$ filter observing 9 out of 10 divertors and neutral pressure measured by a neutral pressure gauge mounted at the pumping slit of the divertor in module 3. Additionally, the graph is extended by the ion saturation current measured by a Langmuir probe located close to the strike line on the surface of the divertor in module 5. Both curves show that increasing neutral pressure is associated with reduced ion flux. High neutral pressure is assured through high recycling of neutrals, which are eventually guided to the pumping gap of the divertor and further to the sub-divertor volume. This is a fundamentally different mechanism for access to detachment than discussed in tokamaks (see for instance \cite{Leonard2018}) with carbon first wall. There, momentum losses along the magnetic flux channel into the divertor are increasing in the high-recycling regime and eventually reduce the parallel particle flux such that the recycling in the divertor is reduced and also neutral pressures decrease \cite{Kallenbach1995, Lipschultz1999}.

Spectroscopic measurements of Balmer-series of hydrogen emission ca. 1~cm above the divertor surface  made in another, but similar, detached discharges (e.g. W7-X \#20181016.009) shows that with increasing line integrated densities a strong increase in electron density above the divertor is observed. Before onset of detachment values of so-called downstream electron density reach $n_{e,d} = 1.2 - 1.4\cdot 10^{20}$ [m$^{-3}$], i.e. are significantly higher than electron densities near separatrix ($n_{e,sep}$ is in the range $4-6\cdot 10^{19}$ [m$^{-3}$]). This difference between upstream and downstream density indicates that divertor operates in high recycling regime. The high recycling regimes (which up to now were only observed on tokamaks) are characterized by a dense divertor plasma, which allows the hydrogenic species in the divertor region to undergo multiple cycles of ionization followed by neutralization on the divertor target plates, before being pumped away. With such high divertor densities we are at the onset of volume recombination, but the amount is still small (ca. 5\%) and other processes are expected to be dominant. Recycling neutrals emitted from the target are released as H$_2$ molecules at wall temperature and trough processes like ionization and charge exchange are kept in the divertor domain. EMC3-Eirene simulations suggest that the charge exchange neutrals undergo a diffusion-like process and at high density form a `diffusion barrier' \cite{Feng2016}.

\section{Summary}
A regime of stable, thermally fully detached island divertor regime with small divertor particle fluxes but still sufficient divertor neutral pressures at Wendelstein 7-X was shown for the first time. Also for the first time in stellarators the high recycling regime was realized, which yielded high neutral concentration in the divertor region. It was possible to establish scenarios with high radiation fraction (at the level of 80-90\%) originating from the carbon ions, low impurity concentration ($Z_{\rm eff} = 1.5$) and high neutral pressure ($p_{\rm n} = 0.07$~Pa), which is compatible with steady state operation of Wendelstein 7-X.  Most importantly discharges with duration of up to 30 seconds could be performed with uncomplicated feedback control. Plasma radiation was a dominant mechanism for dissipation of energy. As the radiation characteristics ($L_z$) of carbon is similar as for nitrogen results obtained here should to a large extent be valid for detachment aided with nitrogen seeding \cite{Effenberg2019}. The results presented in this work form a very promising outlook on the overall steady state compatibility of the detached island divertor concept in future experiments and a stellarator-based reactor.

\bibliographystyle{apsrev} 
\bibliography{prldetachment} 

\end{document}